\documentclass[conference]{IEEEtran}
\IEEEoverridecommandlockouts
\usepackage[sort&compress, square, numbers]{natbib}
\usepackage{amsmath,amssymb,amsfonts}
\usepackage{algorithmic}
\usepackage{graphicx}
\usepackage{textcomp}
\usepackage{xcolor}
\usepackage{multirow}
\usepackage[flushleft]{threeparttable}
\usepackage[normalem]{ulem}
\usepackage{url}

\useunder{\uline}{\ul}{}
\def\BibTeX{{\rm B\kern-.05em{\sc i\kern-.025em b}\kern-.08em
    T\kern-.1667em\lower.7ex\hbox{E}\kern-.125emX}}
\begin{document}

\title{Design of Blockchain-based Travel Rule Compliance System}

\author{
    \IEEEauthorblockN{
    Chaehyeon Lee\IEEEauthorrefmark{1},
    Changhoon Kang\IEEEauthorrefmark{1},
    Wonseok Choi\IEEEauthorrefmark{1},
    Jehoon Lee\IEEEauthorrefmark{1},
    \\Myunghun Cha\IEEEauthorrefmark{2},
    Jongsoo Woo\IEEEauthorrefmark{3},
    and James Won-Ki Hong\IEEEauthorrefmark{1}
}

\IEEEauthorblockA{
    \IEEEauthorrefmark{1}Department of Computer Science and Engineering, POSTECH, Korea
    \\\IEEEauthorrefmark{2}Coinone, Seoul, Korea
    \\\IEEEauthorrefmark{3}Graduate School of Information Technology, POSTECH, Korea
}
}

\IEEEoverridecommandlockouts
\IEEEpubid{\makebox[\columnwidth]{978-1-6654-9538-7/22/\$31.00~\copyright2022 IEEE \hfill} \hspace{\columnsep}\makebox[\columnwidth]{ }}
\maketitle
\IEEEpubidadjcol

\begin{abstract}
In accordance with the guidelines of the Financial Action Task Force (FATF), Virtual Asset Service Providers (VASPs) should comply with a `travel rule', which requires them to exchange originator's and beneficiary's personal information when transferring virtual assets. In this paper, we propose a novel blockchain-based travel rule compliance system that supports fully-decentralized data exchange. The proposed system uses a permissioned blockchain, and thereby eliminates the possibility of leakage of personal information to third parties or even to travel rule service providers, and ensures that travel rule data can be managed securely.
\end{abstract}


\section{Introduction}
\label{sec:intro}

The use of virtual assets for illegal purposes has increased, so, Financial Action Task Force (FATF) proposed guidelines to prevent money laundering and financing of terrorism \citep{fatf2019,fatf20191,fatf20192}. The guidelines include travel rule guidance that aims to have virtual asset service providers (VASP) fulfill their Anti-Money Laundering/Combating the Financing of Terrorism (AML/CFT) obligations \citep{fatf20193}. The travel rule guidance is provided in Recommendation 16 and is intended to restrict terrorists and criminals from moving funds freely by virtual asset transfers \citep{rcmdt}. To this end, the travel rule requires VASPs to share identifying information regarding originator (sender of virtual assets) and beneficiary (receiver of virtual assets). FATF updated the travel rule guidelines in October 2021 \citep{fatf2021}. In response to FATF's guidelines, each country is preparing laws on virtual assets and VASP to comply with FATF regulations. 

In this paper, we propose a novel travel rule compliance system that uses blockchain to comply with the travel rule \citep{pilkington2016blockchain}. In our proposed system, authentication and data exchange between VASPs are performed in a peer-to-peer (P2P) way by using blockchain, so that travel rule data can be exchanged without third-party intermediaries and unnecessary data exposure. Thus, our system is completely decentralized and secure.

\section{Related Work}
\label{sec:rt}
Various companies around the world are developing travel rule compliance systems to comply with the guidelines stipulated by FATF (see Table \ref{tab:table1}). These guidelines have no restrictions on the development of travel rule compliance systems. Most systems \citep{jevans2020travel,sygna,vv,tid,trp} use a central server for authentication or even data-exchange. This may cause a single point of failure and a serious leakage of personal information. To overcome this, blockchain technology can be applied to travel rule solutions. With the use of blockchain technology, it is more efficient than the use of traditional methods to exchange information, and has advantages in various aspects such as cost and storage place management.

OpenVASP \citep{riegelnig2019openvasp} uses Ethereum and does not require authentication processes by third parties. However, blockchain is only used during the process of authenticating the identity; the messaging protocol is used separately from the blockchain in the actual data-exchange process, i.e., OpenVASP only partially uses blockchain.

\begin{table*}[htbp]
\caption{Comparison of Existing Travel Rule Solutions}
\label{tab:table1}
\begin{threeparttable}
\resizebox{\textwidth}{!}{%
\begin{tabular}{|c|c|c|c|c|c|c|}
\hline
\textbf{Service} & \textbf{Company} & \textbf{Decentralized} & \textbf{Blockchain-based} & \textbf{Authentication} & \textbf{Communication} & \textbf{Open-sourced} \\ \hline
TRISA\citep{jevans2020travel}            & CipherTrace      & O                      & X                         & Certificate Authority   & SSL/TLS                & O                     \\ \hline
Sygna Bridge\citep{sygna}     & CoolBitX         & X                      & X                         & Sygna Bridge Server           & SSL/TLS                & X                     \\ \hline
OpenVASP\citep{riegelnig2019openvasp}         & Bitocin Suisse   & O                      & O                         & Ethereum Smart Contract          & Whisper, WAKU, etc.,          & O                     \\ \hline
VerifyVASP\citep{vv}       & Lambda256        & X                      & X                         & Join VerifyVASP Alliance     & HTTPS                  & X                     \\ \hline
Transact ID\citep{tid}      & Netki            & O                      & X                         & Certificate             & HTTPS                  & O                     \\ \hline
TRP\citep{trp}              & ING Group        & O                      & X                         & Direct, Certificate Authority   & RESTful API, HTTP                   & O                    \\ 
\hline
\end{tabular}%
}
\\
\end{threeparttable}
\end{table*}

\section{Design}
Our blockchain-based travel rule compliance system has some advantages. First, thanks to the principle of blockchain, third-party intervention is not required. In some of the existing solutions, the exchange of travel rule data between the VASPs is brokered by a third party, such as the solution-provider's central server. In this case, the customer's personal information held by the exchange may be unnecessarily exposed, and each exchange bears the risk of exposing such personal information. However, by using blockchain, transaction parties can exchange data in a P2P manner, and thereby prevent unwanted exposure from irrelevant third parties.

Travel rule data can be directly exchanged between the two VASPs, and the swapped data can be retained by each to follow decentralization. However, in this case, if one party arbitrarily modifies or deletes data, a discrepancy occurs between the data maintained by the two VASPs. This discrepancy complicates the task of determining which data are genuine and which are not. In contrast, in a blockchain, data cannot be modified or deleted without the consent of the parties involved in the transaction. Even if a VASP succeeds in maliciously modifying the data, the deed cannot be hidden, because all records of data changes remain transparent on the blockchain. Therefore, the travel rule solution that uses blockchain enables secure management of personal information.

\begin{figure} [h]
        \includegraphics[width=0.9\linewidth]{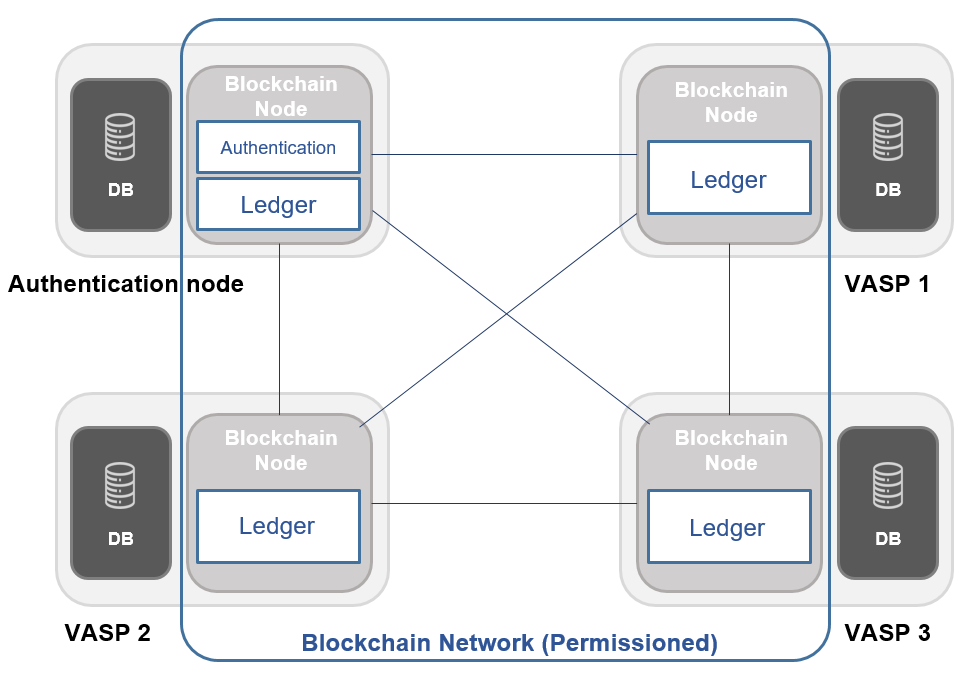}
    \caption{Blockchain-based Travel Rule Compliance System Architecture}
    \label{fig:Fig1}
\end{figure}

The architecture of our travel rule compliance system (see Fig. \ref{fig:Fig1}) takes the form of a permissioned blockchain network to allow only authenticated VASPs to use the system to exchange personal information with counterpart VASPs. A VASP authentication node verifies the identity of the VASP that wants to participate in the network and grants the authority. Authenticated VASPs must operate a single individual node, and must store the data generated in the process of exchanging personal information to comply with the travel rule in its own distributed ledger. Each VASP node retrieves travel rule related data from the connected backend.

To satisfy the travel rule recently specified in Korean law \citep{korealaw}, the travel rule data exchange process is conducted in two steps. When originator at VASP 1 is trying to send virtual asset to beneficiary at VASP 2, the first step is to provide the essential personal information, which includes the name and virtual asset address of originator and beneficiary. After essential personal information is exchanged, virtual assets are transferred, then the personal and transaction data are stored. The second step is to provide additional information. If a previously-executed transfer is determined to be suspicious, VASP 2 can request the real name confirmation information of the originator. This information can be resident registration number, passport number, or alien registration number.

\subsection{Step 1 - Provision of Essential Information}
When a customer of VASP 1 requests virtual asset transfer with beneficiary's virtual asset address, the VASP 1 backend confirms the customer's request and delivers the originator's and beneficiary's virtual asset addresses to the VASP 1 blockchain node. VASP 1 delivers it to VASP 2, then VASP 2 checks whether the originator information is normal. If the originator information does not contain abnormality, then the VASP 2 node brings in the beneficiary's personal information from the backend and transmits it to VASP 1. When the exchange of the originator's and beneficiary's personal information is complete, VASP 1 executes virtual asset remittance and sends transaction information including transaction ID to VASP 2. After confirming the transaction information, VASP 2 informs VASP 1 whether the transaction is normal or not, and if it is normal, confirms the transfer. When a response is received from the VASP 2 node, travel rule data including personal information and transaction information of the originator/beneficiary are recorded in the blockchain ledger by the VASP 1 node.

\subsection{Step 2 - Provision of Additional Information}
The beneficiary VASP may ask the originator VASP to provide the originator's real name confirmation information when fulfilling obligations under the laws, such as Suspicious Transaction Report (STR) \citep{fatf2021}. When suspicious circumstances are found in the transaction executed by Step 1, VASP 2 requests the originator's additional information from VASP 1 by considering the information exchanged in the first step. VASP 1 retrieves the originator's real name information from the backend and delivers it to VASP 2.

\section{Concluding Remarks}

\label{sec:conclusion}
In this paper, we have presented the design of a blockchain-based travel rule compliance system to comply with the FATF's guidance. This allows VASPs to exchange travel rule data in a P2P manner for each transaction. Shared personal information data can be stored and managed securely without unneeded exposure. In future work, we plan to implement this system using Corda enterprise distributed ledger system \citep{corda}. We also plan to link our system with existing \citep{jevans2020travel,sygna,vv,tid,trp,riegelnig2019openvasp} travel rule compliance systems. They are diverse, and the travel rule cannot be applied using a single protocol in the virtual-asset market. This makes expanding our system by integrating with other travel rule systems important.

\section*{Acknowledgments}
This work was supported by CODE and the MSIT(Ministry of Science and ICT), Korea, under the ITRC(Information Technology Research Center) support program(IITP-2022-2017-0-01633) supervised by the IITP(Institute for Information \& Communications Technology Planning \& Evaluation).

\bibliographystyle{unsrt}
\bibliography{mybib}

\end{document}